\documentclass{report}
\usepackage{graphicx,psfrag,amsmath,amssymb,amsfonts,bbm}
\usepackage{svcon2e}

\newcommand{\cc}[1]{{#1}^*}
\newcommand{\scp}[2]{\langle #1 , #2 \rangle}
\newcommand{\ba}{\begin{eqnarray}}
\newcommand{\ea}{\end{eqnarray}}
\newcommand{\bary}{\begin{array}}
\newcommand{\ear}{\end{array}}

\begin{document}
\pagenumbering{arabic} 
\chapter{Classical versus quantum structures: 
the case of pyramidal molecules}

\chapterauthors{
Carlo Presilla\footnotemark[1]\footnotemark[2]\footnotemark[3]\\
Giovanni Jona-Lasinio\footnotemark[1]\footnotemark[2]\\
Cristina Toninelli\footnotemark[1]}

\footnotetext[1]{Dipartimento di Fisica, Universit\`a di Roma ``La Sapienza'',
Piazzale Aldo Moro 2, Roma 00185, Italy}
\footnotetext[2]{Istituto Nazionale di Fisica Nucleare, 
Sezione di Roma 1, Roma 00185, Italy}
\footnotetext[3]{Istituto Nazionale per la Fisica della Materia, 
Unit\`a di Roma 1 and Center for Statistical Mechanics and Complexity, 
Roma 00185, Italy}

\begin{abstract} 
In a previous paper we proposed  
a model to describe a gas of pyramidal molecules interacting 
via dipole-dipole interactions. 
The interaction  modifies the tunneling properties between the classical 
equilibrium configurations of the single molecule and, 
for sufficiently high pressure, the molecules
become localized in these classical configurations. 
The model explains quantitatively the shift to zero-frequency of the 
inversion line observed upon increase of the 
pressure in a gas of ammonia or deuterated ammonia. 
Here we analyze further the model especially with respect to 
stability questions.
\end{abstract}  

\section{Introduction} 
\label{starting}

The behavior of gases of pyramidal molecules, 
i.e. molecules of the kind $XY_3$ like ammonia $NH_3$,
has been the object of investigations since the early developments
of quantum mechanics \cite{Hund}.
In recent times the problem has been discussed again in several papers
\cite{jonaclaverie,jonaclaverie2,Wightman,gm,jptprl,jpt}
from a stationary point of view while in \cite{gms,gspra,gspreprint}
a dynamical approach has been attempted.
For a short historical sketch of the issues involved we refer to 
\cite{jonaclaverie,Wightman,jptprl}.

In \cite{jptprl} we have constructed a simplified mean-field 
model of a gas of pyramidal molecules which allows a direct comparison 
with experimental data. 
Our model predicts, for sufficiently high inter-molecular interactions, 
the presence of two degenerate ground states corresponding to the  
different localizations of the molecules. 
This transition to localized states gives a reasonable 
explanation of the experimental results 
\cite{BL1, BL2, Birnbaum}. 
In particular, it describes quantitatively, without free parameters, 
the shift to zero-frequency of the inversion line of $NH_3$ and $ND_3$ 
on increasing the pressure.

In the present paper we first reconsider our model from the
stand point of stationary many-body theory clarifying the meaning
of the mean-field energy levels. 
We then analyze the mean field states with respect to the energetic
stability. 
The conclusions agree with those found in \cite{gspreprint} via a
dynamical analysis of the same type of model to which dissipation is
added.

\section{The model} 
\label{model}  

We model the gas as a set of two-level quantum systems, 
that mimic the inversion degree of freedom of an isolated molecule,
mutually interacting via the dipole-dipole electric force.

The Hamiltonian for the single isolated molecule is assumed of the form
$-\frac{\Delta E}{2}\sigma^x$,
where $\sigma^x$ is the Pauli matrix with symmetric and antisymmetric 
delocalized tunneling eigenstates $\varphi_+$ and $\varphi_-$
\begin{equation}
\sigma^x =
\left( \begin{array}{cc} \phantom{-}0&1\\-1&0 \end {array} \right)
\qquad
\varphi_+ = \frac{1}{\sqrt 2}
\left( \begin{array}{c} 1 \\ 1 \end {array} \right) 
\qquad
\varphi_- = \frac{1}{\sqrt 2} 
\left( \begin{array}{c} \phantom{-}1 \\ -1 \end {array} \right).
\label{12}
\end{equation}
Since the rotational degrees of freedom of the single pyramidal molecule 
are faster than the inversion ones, 
on the time scales of the inversion dynamics set by $\Delta E$ the molecules 
feel an effective attraction arising from the angle averaging of the 
dipole-dipole interaction at the temperature of the experiment 
\cite{Keesom}.
The localizing effect of the dipole-dipole interaction between two
molecules $i$ and $j$ can be represented by an interaction term 
of the form $-g_{ij} \sigma^z_i \sigma^z_j$, with $g_{ij}>0$,
where $\sigma^z$ is the Pauli matrix with left and right localized 
eigenstates $\varphi_L$ and $\varphi_R$ 
\begin{equation}
\sigma^z =
\left( \begin{array}{cc} 1&\phantom{-}0\\0&-1 \end {array} \right)
\qquad
\varphi_L = 
\left( \begin{array}{c} 1 \\ 0 \end {array} \right) 
\qquad
\varphi_R = 
\left( \begin{array}{c} 0 \\ 1 \end {array} \right).
\label{LR}
\end{equation} 
The Hamiltonian for $N$ interacting molecules then reads 
\begin{eqnarray}
H&=& -\frac{ \Delta E}{2} \sum_{i=1}^N
1_1\otimes 1_2\otimes\ldots\otimes\sigma^x_i\otimes\ldots 1_N
\nonumber \\ &&
- \sum_{i=1}^{N} \sum_{j=i+1} ^{N} g_{ij} \,
1_1 \otimes \ldots \otimes \sigma^z_i \otimes \ldots 
\otimes \sigma^z_j \otimes \ldots \otimes 1_N.
\label{HN}
\end{eqnarray}
For a gas of moderate density, we approximate 
the behavior of the $N \gg 1$ molecules with the mean-field Hamiltonian
\begin{equation}
h[\psi]=-\frac{\Delta E}{2}\sigma^x-
G\scp{\psi}{\sigma^z \psi} \sigma^z,
\label{acca}
\end{equation}
where $\psi$ is the single-molecule state ($\scp{\psi}{\psi}=1$)
to be determined
self-consistently by solving the nonlinear eigenvalue problem
associated to (\ref{acca}).
The parameter $G$ represents the dipole interaction energy of a 
single molecule with the rest of the gas.
This must be identified with a sum over all possible molecular 
distances and all possible dipole orientations calculated with the 
Boltzmann factor at temperature $T$.
Assuming that the equation of state for an ideal gas applies, 
we find \cite{jptprl}
\begin{equation}
G=\frac{4 \pi}{9} \left( \frac{T_0}{T}\right)^2 P d^3,
\label{G}
\end{equation}
where $T_0= \mu^2/(4\pi \varepsilon_0\varepsilon_r d^3 k_B)$,
$\varepsilon_0$ and $\varepsilon_r$ being the vacuum and relative 
dielectric constants, $d$ the molecular collision diameter and 
$\mu$ the molecular electric dipole moment.
Note that, at fixed temperature, the mean-field interaction constant $G$ 
increases linearly with the gas pressure $P$.

\section{Molecular states} 
\label{states}  
The nonlinear eigenvalue problem associated to (\ref{acca}), namely
\begin{equation}
h[\psi_\mu]\psi_\mu = \mu \psi_\mu
\qquad \scp{\psi_\mu}{\psi_\mu}=1,
\end{equation} 
has different solutions depending on the value of the ratio $G/ \Delta E$.
If $G/\Delta E<\frac12$, we have only two solutions corresponding to the 
delocalized eigenstates of an isolated molecule
\begin{eqnarray}
&&\psi_{\mu_1}=\varphi_+ \qquad \mu_1=-\Delta E/2   \label{s1} \\ 
&&\psi_{\mu_2}=\varphi_- \qquad \mu_2=+\Delta E/2 . \label{s2}
\end{eqnarray}
If $G/ \Delta E>\frac12$, there appear also two new solutions
\begin{eqnarray}
&&\psi_{\mu_3}=
\sqrt{\frac{1}{2} +\frac{\Delta E}{4G}}~\varphi_+
+\sqrt{\frac{1}{2}-\frac{\Delta E}{4G}}~\varphi_-
\qquad \mu_3=-G \label{s3}\\
&&\psi_{\mu_4}=
\sqrt{\frac{1}{2} +\frac{\Delta E}{4G}}~\varphi_+
-\sqrt{\frac{1}{2}-\frac{\Delta E}{4G}}~\varphi_-
\qquad \mu_4=-G \label{s4}
\end{eqnarray}
which in the limit $G\gg \Delta E$ approach the localized states
$\varphi_L$ and $\varphi_R$, respectively. 
Solutions (\ref{s3}) and (\ref{s4}) are termed chiral in the sense that 
$\psi_{\mu_4} = \sigma^x \psi_{\mu_3}$. 

\begin{figure}
\centering
\psfrag{G/DE}[t][][0.9]{$G/\Delta E$}
\psfrag{E}[b][][0.9]{$e_i/\Delta E$}
\includegraphics[width=0.6\textwidth,clip]{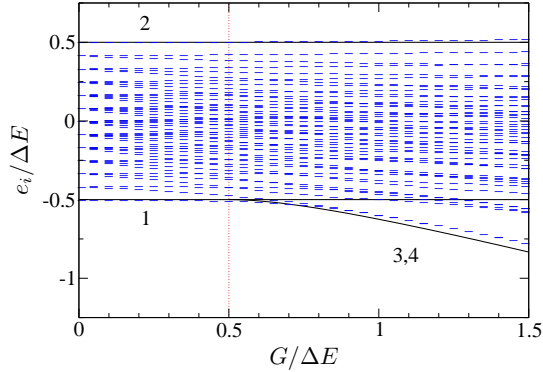}
\caption{Single-molecule energies $e_i$ (solid lines) of the four stationary 
states $\psi_{\mu_i}$, $i=1,2,3,4$, as a function of the ratio $G/\Delta E$.
The dashed lines are the eigenvalues, divided by $N$, of the Hamiltonian
(\ref{HN}) with $g_{ij}=G/N$ and $N=12$.} 
\label{energy}
\end{figure}

The states $\psi_\mu$ determined above,
are the stationary solutions $\psi(t) = \exp(i\mu t/ \hbar) \psi_\mu$ 
of the time-dependent nonlinear Schr\"odinger equation
\begin{equation}
i \hbar \frac{\partial}{\partial t} \psi(t) = h[\psi] \psi(t).
\label{tdse}
\end{equation}
The generic state $\psi(t)$ solution of this equation 
has an associated conserved energy given by
\begin{eqnarray}
\mathcal{E}[\psi] =
-\frac{\Delta E}{2} \scp{\psi}{\sigma^x \psi}
-\frac{G}{2} \scp{\psi}{\sigma^z \psi}^2,
\end{eqnarray}
The value of this functional calculated at the stationary solutions
(\ref{s1}-\ref{s4}) provides the corresponding single-molecule energies 
$e_i=\mathcal{E}[\psi_{\mu_i}]$
\begin{eqnarray}
&&e_1=-\Delta E/2 
\nonumber\\
&&e_2= +\Delta E/2 
\\ 
&&e_3=e_4= -\frac{\Delta E}{2}-\frac{1}{2G}
\left( \frac{\Delta E}{2}-G \right)^2.
\nonumber
\end{eqnarray}
These energies are plotted in Fig. \ref{energy} as a function of the 
ratio $G/\Delta E$.
The state effectively 
assumed by the molecules in the gas will be that 
with the minimal energy, namely the symmetric delocalized state $\psi_{\mu_1}$ 
for $G/\Delta E< \frac12$ or one of the two degenerate chiral states 
for $G/\Delta E> \frac12$.

The above results imply a bifurcation of the mean-field ground state 
at a critical interaction strength $G=\Delta E/2$. 
According to Eq. (\ref{G}), this transition can be obtained
for a given molecular species 
by increasing the gas pressure above the critical value 
\begin{equation}
P_\mathrm{cr}=\frac{9}{8 \pi} P_0 \left(\frac{T}{T_0}\right)^2,
\label{pcr}
\end{equation}
where $P_0=\Delta E/d^3$.
In Table \ref{p0t0} we report the values of $T_0$ and $P_0$ calculated for
different pyramidal molecules.
\begin{table}
\begin{center}
\begin{tabular}{lccccc}
\hline
\hline
&$\Delta E$ (cm$^{-1}$)&$\mu$ (Debye)&$d$ (\AA)&$T_0$ (Kelvin)&$P_0$ (atm)\\
\hline
$NH_3$&0.81&1.47&4.32&193.4&1.97\\
$ND_3$&0.053&1.47&4.32&193.4&0.13\\
$PH_3$&$3.34\times 10^{-14}$&0.57&--&29.1&8.11$\times~10^{-14}$\\
$AsH_3$&$2.65\times 10^{-18}$&0.22&--&4.3&6.44$\times~10^{-18}$\\
\hline
\hline
\end{tabular}
\end{center}
\caption{
Measured energy splitting $\Delta E$, 
collision diameter $d$, and 
electric dipole moment $\mu$, 
for different pyramidal molecules as taken from \cite{Townes,HCP}. 
In the fourth and fifth columns we report the temperature $T_0$ 
and the pressure $P_0$ evaluated as described in the text. 
In the case of $PH_3$ and $AsH_3$ the collision diameter, not available, 
is assumed equal to that measured for $NH_3$ and $ND_3$. 
We used $\varepsilon_r=1$.}
\label{p0t0}      
\end{table}

\section{Inversion line}

\begin{figure}
\centering   
\psfrag{G/DE}[t][][0.9]{$G/\Delta E$}
\psfrag{hnu/DE}[b][][0.9]{$h \bar{\nu}/\Delta E$}
\includegraphics[width=0.6\textwidth,clip]{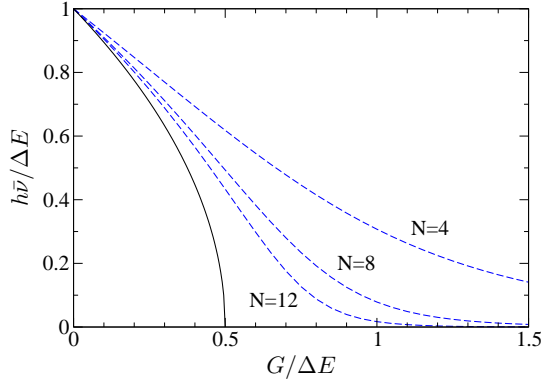}
\caption{Inversion line frequency as a function of the ratio $G/\Delta E$
in the mean field model (solid line) and obtained from the 
Hamiltonian (\ref{HN}) with $g_{ij}=G/N$ and $N=4$, 8, and 12.} 
\label{gap}
\end{figure}

When a gas of pyramidal molecules which are in the delocalized ground state
is exposed to an electromagnetic radiation of angular frequency 
$\omega_0 \sim \Delta E /\hbar$, 
some molecules can be excited from the state $\varphi_+$
to the state $\varphi_-$. 
For a non-interacting gas this would imply the presence 
in the absorption or emission spectrum of an inversion
line of frequency $\bar{\nu}=\Delta E /h$.
Due to the attractive dipole-dipole interaction, the value of
$h\bar{\nu}$ evaluated as the energy gap between the many-body first 
excited level and the ground state is decreased with respect to the 
noninteracting case by an amount of the order of $G$.
As shown in Fig. (\ref{gap}), the value of the inversion line frequency is 
actually a function of the number $N$ of molecules and
in the limit $N \gg 1$ approaches the mean field value \cite{jptprl}   
\begin{equation}
\bar{\nu}=
\frac{\Delta E}{h}\left( 1-\frac{2G}{\Delta E}\right)^{\frac12}.
\label{freq}
\end{equation}
According to (\ref{freq}), the inversion line is obtained only in 
the range $0 \leq G \leq \Delta E/2$ and its frequency vanishes
at $G=\Delta E/2$.

In \cite{jptprl} we have compared the mean field theoretical prediction 
for the inversion line with the spectroscopic data available for 
ammonia \cite{BL1,BL2} and deuterated ammonia \cite{Birnbaum}.
In these experiments the absorption coefficient of a cell
containing $NH_3$ or $ND_3$ gas at room temperature   
was measured at different pressures, i.e. according to (\ref{G}) at 
different values of the interaction strength $G$.
The measured frequency $\bar{\nu}$ decreases by increasing 
$P$ and vanishes for pressures greater than a critical value.
This behavior is very well accounted for by the the mean field 
prediction (\ref{freq}).
In particular, the critical pressure evaluated according to 
Eq. (\ref{pcr}) 
with no free parameters, at $T=300$ K is
$P_\mathrm{cr} = 1.695$ atm for $NH_3$ 
and $P_\mathrm{cr} = 0.111$ atm for $ND_3$ is in very good agreement 
with the experimental data.

\section{Energetic stability of the molecular states} 
\label{stability}  

In order to discuss the energetic stability of the mean field molecular
states found in Section \ref{states} we introduce the free energy 
$\mathcal{F}[\psi] = \mathcal{E}[\psi] - \mu \scp{\psi}{\psi}$.
The stationary solutions of Eq. (\ref{tdse}) then can be viewed as the 
critical points of the Hamiltonian dynamical system
\begin{equation}
\label{hamilton}
i \hbar \frac{\partial}{\partial t} 
\left( \begin{array}{c} \psi\phantom{^*} 
\\ \cc{\psi} \end {array} \right) 
= 
\left( \begin{array}{cc} 
\phantom{-}0 & 1 \\ 
-1 & 0 
\end{array}\right)
\left(\begin{array}{c} 
\frac{\delta\mathcal{F}[\psi]}{\delta\psi} 
\\ 
\frac{\delta\mathcal{F}[\psi]}{\delta\cc{\psi}}  
\end {array}\right) .
\end{equation}
Under the effect of a perturbation which dissipates energy,
a stationary state $\psi_\mu$ will remain stable only if 
$\mathcal{F}[\psi_\mu]$ is a minimum. 
Therefore we are interested in exploring the nature of the
extremal values $\mathcal{F}[\psi_\mu]$ of the free energy functional.
In general, this can be done in terms of the eigenvalues and the 
eigenvectors of the linearization matrix associated to the
dynamical system (\ref{hamilton}) as explained in \cite{dap}
for a Gross-Pitaevskii equation.
Here, due to the simplicity of the model, we can provide a more direct 
analysis.
   
For a variation of the stationary solution 
$\psi_\mu \to \psi_\mu + \delta\phi$,
up to the second order in $\delta\phi$, we have
\begin{equation}
\mathcal{F}[\psi_\mu + \delta\phi]
= \mathcal{F}[\psi_\mu] + \delta^2\mathcal{F}[\psi_\mu,\delta\phi],
\end{equation} 
where
\begin{eqnarray}
\delta^2\mathcal{F}[\psi_\mu,\delta\phi] &=&
-\frac{\Delta E}{2} \scp{\delta\phi}{\sigma^x \delta\phi}
-\mu \scp{\delta\phi}{\delta\phi}
\nonumber\\&&
-G \left( 
  \scp{\psi_\mu}{\sigma^z \psi_\mu} 
  \scp{\delta\phi}{\sigma^z \delta\phi}
+ \scp{\psi_\mu}{\sigma^z \delta\phi} 
  \scp{\delta\phi}{\sigma^z \psi_\mu}
\phantom{\frac12}\right.\nonumber\\&&\left.\qquad
+\frac{1}{2}\scp{\psi_\mu}{\sigma^z \delta\phi}^2
+\frac{1}{2}\scp{\delta\phi}{\sigma^z \psi_\mu}^2
\right).
\end{eqnarray}
The variation $\delta\phi$ can be taken in the most general form 
\begin{equation}
\delta\phi = ae^{i\theta_a} \varphi_+ + be^{i\theta_b} \varphi_-,
\end{equation}
where $\varphi_\pm$ are the delocalized tunneling eigenstates (\ref{12})
and $a$, $b$, $\theta_a$, and $\theta_b$ arbitrary real 
parameters with the constraint that 
$\scp{\psi_\mu + \delta\phi}{\psi_\mu + \delta\phi}=1
+\mathcal{O}(\delta\phi^2)$.
By writing 
\begin{equation}
\psi_\mu = a_\mu \varphi_+ + b_\mu \varphi_-,
\end{equation}
with the real coefficients $a_\mu$ and $b_\mu$ deduced by 
Eqs. (\ref{s1}-\ref{s4}), the above constraint implies 
\begin{equation}
a_\mu a \cos\theta_a + b_\mu b \cos\theta_b =0.
\label{condition}
\end{equation}
The second variation of the free energy evaluated at
the four stationary solutions $\psi_{\mu_i}$, $i=1,2,3,4$,
with the condition (\ref{condition}) gives
\begin{eqnarray}
\delta^2\mathcal{F}[\psi_{\mu_1}] &=&
G~ 2b^2 \left( \frac{\Delta E}{2G} - \cos^2 \theta_b \right),
\\
\delta^2\mathcal{F}[\psi_{\mu_2}] &=&
G~ 2a^2 \left( -\frac{\Delta E}{2G} - \cos^2 \theta_a \right),
\\
\delta^2\mathcal{F}[ \psi_{\mu_k}] &=&
G~ \Biggl[ 
  2a^2 \left(1+\frac{\Delta E}{2G} \right) \cos^2\theta_a 
+ 2b^2 \left(1-\frac{\Delta E}{2G} \right) \cos^2\theta_b 
\nonumber\\ && \qquad +
\left( a\sqrt{1-\frac{\Delta E}{2G}}\sin\theta_a \mp
       b\sqrt{1+\frac{\Delta E}{2G}}\sin\theta_b \right)^2
\Biggr], \qquad
\end{eqnarray}
where $k=3,4$ and the signs $\mp$ refer respectively to $k=3$ and 
$k=4$.
We see that, for the state $\psi_{\mu_1}$, the variation 
$\delta^2\mathcal{F}$ is always positive for $G< \frac{\Delta E}{2}$ 
and can be negative for $G > \frac{\Delta E}{2}$. 
The variation $\delta^2\mathcal{F}$ is always negative
in the case of $\psi_{\mu_2}$.
For the states $\psi_{\mu_3}$ and $\psi_{\mu_4}$, 
which exist only for $G > \frac{\Delta E}{2}$, 
the variation $\delta^2\mathcal{F}$ is always positive.
We conclude that the free energy has a single minimum 
in correspondence of the delocalized state $\psi_{\mu_1}$ 
when $G< \frac{\Delta E}{2}$,
and two degenerate minima 
in correspondence of the chiral states $\psi_{\mu_3}$ and 
$\psi_{\mu_4}$ when $G> \frac{\Delta E}{2}$.
The energetic stability analysis is summarized in Table \ref{esa}.
Note that our results coincide with those reported in \cite{gspreprint} 
where a standard linear stability analysis is performed for the same 
model considered here to which an explicit norm-conserving dissipation 
is added. 
\begin{table}
\centering
\begin{tabular}{lcccc}
\hline
\hline
& $\psi_{\mu_1}$ & $\psi_{\mu_2}$ & $\psi_{\mu_3}$ & $\psi_{\mu_4}$ \\
\hline
$\mathcal{F}[\psi_\mu]$ & $-\Delta E$ & $+\Delta E$ & 
$\frac{G}{2} \left[1- \left(\frac{\Delta E}{2G}\right)^2 \right]$ &  
$\frac{G}{2} \left[1- \left(\frac{\Delta E}{2G}\right)^2 \right]$ \\
$G< \frac{\Delta E}{2}$ & $\delta^2\mathcal{F}>0$ & $\delta^2\mathcal{F}<0$ & & \\
& minimum & maximum & & \\
$G> \frac{\Delta E}{2}$ & $\delta^2\mathcal{F}\gtrless 0$ & $\delta^2\mathcal{F}<0$ & $\delta^2\mathcal{F}>0$ & $\delta^2\mathcal{F}>0$ \\
& saddle point & maximum & minimum & minimum\\
\hline
\hline
\end{tabular}
\caption{
Value of the free energy $\mathcal{F}[\psi_\mu]$ and sign of its second 
variation at the four extrema $\psi_{\mu_i}$, $i=1,2,3,4$.}
\label{esa}      
\end{table}

\section{Conclusions}

The specific prediction of our model for the critical pressure
$P_\mathrm{cr}$ in terms of the electric dipole $\mu$ of the molecule, 
its size $d$, the splitting $\Delta E$ and the temperature $T$ 
of the gas, successfully verified in the case of ammonia,  
could be experimentally tested also for other pyramidal gases
for which,  Eqs. (\ref{pcr}) and (\ref{freq}) predict the scaling law
\begin{equation}
\frac{\bar\nu_{XY_3}(P)}{\bar\nu_{XY_3}(0)}=
\frac{\bar\nu_{X'Y'_3}(\gamma P)}{\bar\nu_{X'Y'_3}(0)},
\label{pgamma}
\end{equation}
where
$\gamma = \left. P_\mathrm{cr}\right._{X'Y'_3}/
\left.P_\mathrm{cr}\right._{XY_3}$.

Our model applies not only to molecules $XY_3$ but also to 
their substituted derivatives $XYWZ$.
An important difference between the two cases is that for 
$XY_3$ the localized states can be obtained one from the other either 
by rotation or by space inversion, while for $XYWZ$ they can be 
connected only by space inversion. 
This implies that $XYWZ$ molecules at a pressure greater 
than the critical value are chiral and therefore optically active.
The measurement of the optical activity of pyramidal gases 
for $P>P_\mathrm{cr}$ would allow a direct verification of this
prediction.

\end{document}